\documentclass[12pt]{article}
\usepackage{verbatim}
\usepackage{amsfonts}
\usepackage{graphics}
\usepackage{amsmath}
\usepackage{times}
\usepackage{appendix}
\usepackage{color}
\usepackage{enumerate}
\usepackage{fancyhdr,latexsym,amsmath,amsfonts,amssymb,amsbsy,amsthm,url}
\usepackage[margin=0.5in,footskip=0.25in]{geometry}
\usepackage{graphics,graphicx,epsfig}
\usepackage{breqn}
\usepackage{caption,subcaption,float,subfloat}
\usepackage{pdflscape}
\usepackage{hyperref}
\usepackage[ruled,vlined]{algorithm2e}

\newtheorem{assumption}{Assumption}

\usepackage[english]{babel}
\usepackage[dvipsnames]{xcolor}
\makeatletter

\renewcommand{\section}{
	\@startsection
	{section}
	{1}
	{0pt}
	{1.1\baselineskip}
	{0.2\baselineskip}
	{\sc \centering}
}

\renewcommand{\subsection}{
	\@startsection
	{subsection}
	{1}
	{0pt}
	{1.1\baselineskip}
	{0.2\baselineskip}
	{\sc \centering}
}

\renewcommand{\subsubsection}{
	\@startsection
	{subsubsection}
	{1}
	{0pt}
	{1.1\baselineskip}
	{0.2\baselineskip}
	{\sc \centering}
}

\makeatother

\usepackage[flushleft]{threeparttable}
\usepackage{rotating,booktabs,multirow}
\usepackage{colortbl}
\usepackage{makecell,cellspace,caption}

\begin{document}
	
\title{\large\sc Single Event Transition Risk: A Measure for Long Term Carbon Exposure}
\normalsize
\author{\sc{Suryadeepto Nag} \thanks{Indian Institute of Science Education and Research Pune, Pune-411008, Maharashtra, India, e-mail: suryadeepto.nag@students.iiserpune.ac.in}
\and \sc{Siddhartha P. Chakrabarty} \thanks{Department of Mathematics and School of Business, Indian Institute of Technology Guwahati, Guwahati-781039, Assam, India, e-mail: pratim@iitg.ac.in,
Phone: +91-361-2582606, Fax: +91-361-2582649}\and \sc{Sankarshan Basu} \thanks{Department of Finance and Accounting, Indian Institute of Management Bangalore, Bengaluru-560076, Karnataka, India, e-mail: sankarshan.basu@iimb.ac.in}}
\date{}
\maketitle
\begin{abstract}
Although there is a growing consensus that a low-carbon transition will be necessary to mitigate the accelerated climate change, the magnitude of transition-risk for investors is difficult to measure exactly. Investors are therefore constrained by the unavailability of suitable measures to quantify the magnitude of the risk and are forced to use the likes of absolute emissions data or ESG scores in order to manage their portfolios. In this article, we define the Single Event Transition Risk (SETR) and illustrate how it can be used to approximate the magnitude of the total exposure of the price of a share to low-carbon transition. We also discuss potential applications of the single event framework and the SETR as a risk measure and discuss future direction on how this can be extended to a system with multiple transition events. 
\end{abstract}

\section{Introduction}

Growing climate consciousness has led to an increased interest in various low-carbon transition policies and scenarios. Social responsibilities aside, to be long on carbon intensive stocks during such a transition can prove to be rather risky for investors, and therefore investors must factor in low-carbon transition risk into their portfolios. Currently, the quantities that an investor can use to hedge their portfolio are limited, and therefore they have to rely on simple quantities like absolute carbon emissions of firms, emissions intensities or quantities derived from composite scores like Environmental, Social and Corporate Governance (ESG) scores \cite{madhavan2021toward} or Brown-Green Scores \cite{gorgen2020carbon}. Emissions, emissions intensity or ESG reports are usually published by firms under voluntary disclosures, as a part of Corporate Social Responsibility (CSR). These quantities are readily available, making them convenient tools for assessing carbon risks associated with firms. While there is existing literature on portfolio optimization strategies using absolute emissions \cite{andersson2016hedging}, emissions intensity \cite{fang2019sustainable} Brown-Minus-Green (BMG) risk factor \cite{roncalli2020measuring,roncalli2021market} and ESG scores \cite{baker2018financing,pastor2021sustainable,pedersen2021responsible} it is imperative to note that these quantities only convey a relative degree of risk between countries and do not quantify or give an approximation of the actual exposure. For instance, an ESG score tells an investor that one firm's  production is cleaner than another's, and the former with lesser emissions is likely to have lesser transition risk than the latter. However, it does not give the investor any information about the magnitude of the risk, or even the relationship between the two risks. For instance, just because one firm's emissions are twice as much as another firm's, it does not necessarily mean that the risk too, will be twice as much. Furthermore, the relationship between ESG scores and financial performance is not consistent among different studies done with different sets of data. For instance, a study on firms in Latin America find a negative association between ESG scores and financial performance \cite{duque2021environmental}. This, however, disagrees with research on S\&P500 firms which finds the impact to be time dependent, \textit{i.e.,} a negative association is found on the short term but a positive association is found for the long term \cite{nollet2016corporate}. Other studies find more fundamental inconsistencies in the relationship between ESG scores and financial performance, where market-based and accounting-based financial performance were observed to vary differently with ESG scores for firms in Germany, where a positive association was found with accounting-based performance but no association was found for market-based financial performance \cite{velte2017does}. 
	
Therefore, the existing measures are at best, proxy measures that help in qualitative comprehension of relative risk between stocks but fail to give insights about the actual exposure of the stock to transition risk. Therefore, despite rich literature on sustainable investing, the models and strategies are limited due to the unavailability of a more direct measure of the transition risk. This problem would be fixed in the presence of a methodology that allowed investors and analysts to measure the exposure of the risk from empirical facts like emissions or emissions intensity, or an approximation of the same. Given that when and how low-carbon transition will take place is likely to vary from place to place and also be influenced by a plethora of events and processes, those occurring presently and those yet to occur, hence, determining an exact measure of the exposure is an ambitious and futile exercise. Nevertheless, it would be useful to make some simplifying assumptions and develop an approximate measure for the exposure. Investors could then use this approximate measure to hedge their portfolios and divest from fossil fuel intensive firms. Accordingly, we propose for one such measure, namely, the Single Event Transition Risk (SETR).
	
The SETR is a novel quantity developed in the context of the thought experiment that asks what would be the premium-risk relationship if carbon-transition was triggered by a single event. It is a hypothetical quantity estimated based on risk-premiums estimated in the real market, and some ideal market assumptions. Although the SETR is calculated under ideal assumptions, we expect its utility in portfolio management and risk analysis will be significant. This is akin to ``fatal shock models'' \cite{marshall1967multivariate,gut2005realistic} in the reliability literature which have been used in the context of credit risk and insurance modeling \cite{lindskog2003common}. 
	
In this article, we develop the idea of a single event framework and propose the SETR as a proxy measure for risk, and argue why this is a useful measure to have, when compared to existing quantifiers used in the context of carbon transition risk. We also highlight future direction in the context of the utility and applicability of the single event framework and the SETR. Accordingly, We define the SETR as the total exposure of a share of a firm at any point in time, if the entire loss of value due to a low-carbon transition is incorporated at that time. The single event is a simplification of the real scenario where the total risk is likely to be triggered by a series of events rather than a single one.

\section{The Single Event Transition Risk}

\subsection{Definition}

The SETR is defined in the context of a single event framework, where the transition risk is incorporated into the value of the stock due to a single ``fatal'' shock. Fatal shocks are defined as shocks which are stronger than some critical level, after which a component of the system is entirely killed \cite{marshall1967multivariate}. This is analogous to our system where a single shock kills the carbon-associated component of the share value. This is equivalent to considering a hypothetical market, which is identical to real markets in all ways except that in this market, the low-carbon transition happens due to a single shock.
\begin{figure}[H]
\begin{subfigure}{.5\textwidth}
\centering
\includegraphics[width=1\linewidth]{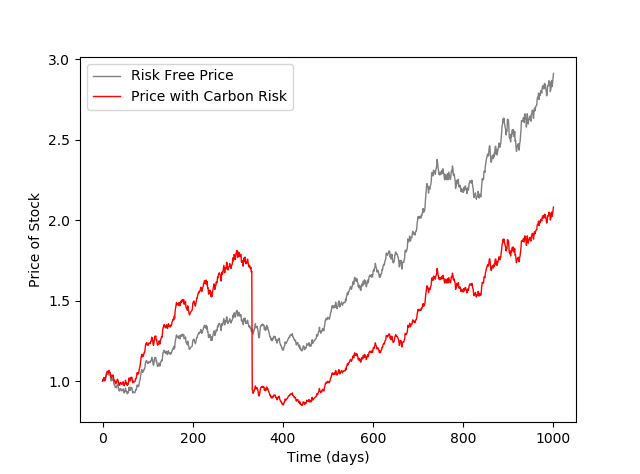}
\caption{$t=331$}
\label{fig:sub-third}
\end{subfigure}
\begin{subfigure}{.5\textwidth}
\centering
\includegraphics[width=1\linewidth]{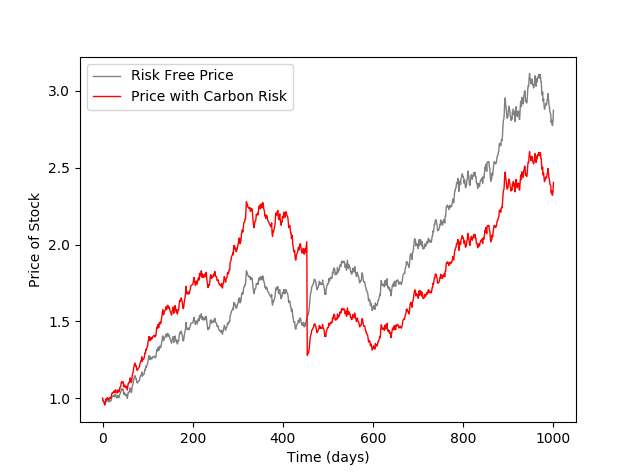}
\caption{$t=453$}
\label{fig:sub-second}
\end{subfigure}
\newline
\begin{subfigure}{.5\textwidth}
\centering
\includegraphics[width=1\linewidth]{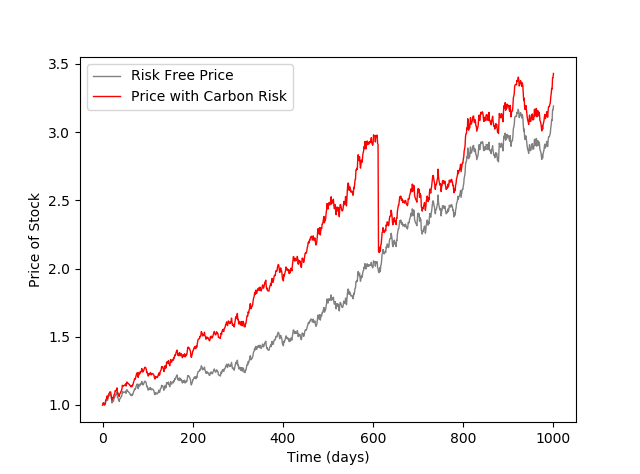}
\caption{$t=611$}
\label{fig:sub-ninth}
\end{subfigure}
\begin{subfigure}{.5\textwidth}
\centering
\includegraphics[width=1\linewidth]{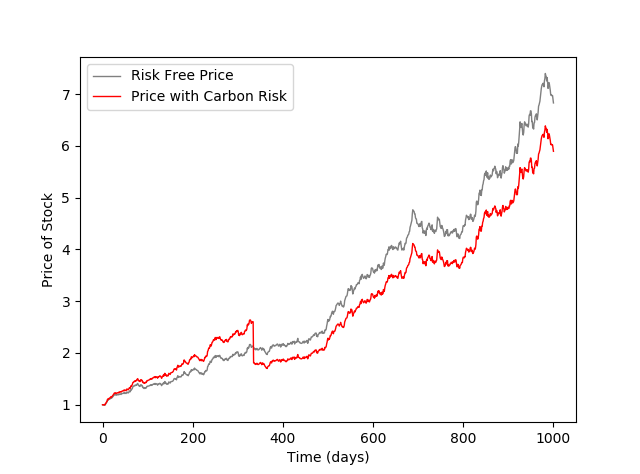}
\caption{$t=334$}
\label{fig:sub-fourth}
\end{subfigure}
\caption{\textbf{Illustrations of transition risk in a single event framework}: The time series of the price of a risk free stock has been contrasted with a corresponding carbon intensive stock with a constant carbon premium and a SETR. The stock price has been simulated using a Geometric Brownian Motion with $\mu=0.0015$ and $\sigma=0.01$. The constant risk premium is set at $0.001$ per day and the arrival process has been assumed to be exponential with scale equal to $750$ days. Four realizations have been plotted with time of transitions (a) $t=331$ (b) $t=453$ (c) $t=611$ and (d) $t=334$.}
\label{fig:Simulations}
\end{figure}
Nevertheless, the single event scenario gives a worst-case estimate and therefore finds utility in portfolio considerations. An important point to note is that the SETR does not only quantify the value of the risk for a transition to a zero-carbon level. In fact, the framework can be used for a transition to any given carbon level, as long as the market is not expected to transition any more after that. In spite of the SETR being estimated under hypothetical single event considerations, there are several real scenarios which would correspond almost exactly to a single event risk. For instance, unexpected government regulation on carbon emissions of firms. If the government of a country passes legislation restricting the emissions of a firms, one would expect the value of the firm to take a large hit based on the loss of productivity due to the newly introduced carbon-cap. Most of this loss of share value would be incorporated in a small interval of time. Measures like carbon taxes which are expected to force firms to transition in more passive ways do not correspond exactly to the single event framework, but this only demands for models which can translate from carbon tax rates to firm revenue and profits, and therefore, for a given tax rates, if it is possible to calculate the shortfall of profits and revenue and other financials, it is possible to quantify the magnitude of the SETR. In other types of government legislation like carbon rationing or carbon allowances, where firms are allowed to emit a maximum amount of emissions (which may even be firm or sector specific), the SETR is nearly an exact model. In the subsequent sections, we will develop a method where we can calculate the SETR for firms based on price movements and financials, and also demonstrate its applications to pricing risks in the paradigm of firm-level carbon allowance legislation.
	
\subsection{The Carbon Risk Premium}

A major breakthrough in the climate finance literature has been the discovery of a significant carbon premium in multiple contexts across different countries. The carbon premium can be defined as the abnormal excess return on a stock or a portfolio, attributed entirely to the carbon(or other greenhouse gas) emissions of the firm or set of firms in the context of a portfolio. The carbon premium has been estimated in several empirical studies, both at the firm level \cite{bolton2021investors,bolton2021global} and portfolio level \cite{oestreich2015carbon}. Since we are interested in estimating the exposure for a firm, we concern ourselves with the firm level carbon premium. The firm level carbon premium is an artefact of price patterns in financial markets, where firms with disproportionately large carbon emissions earn returns higher than one would expect for a firm with the same financials but lower emissions. This premium can be thought of as an incentives for investors to buy or hold shares of ``dirty'' (carbon intensive) firms. The incentive is necessary in order to offset the higher transition risk present with carbon intensive firms. In the absence of such an incentive, a rational trader would sell the shares of dirty firms as they come with a risk, but no additional benefit. We use this heuristic to arrive at an appropriate relationship between a correctly priced carbon premium and a SETR. 
	
\subsection{From Carbon Premium to Carbon Risk}
\begin{figure}[h]
\centering
\includegraphics[width=1\linewidth]{"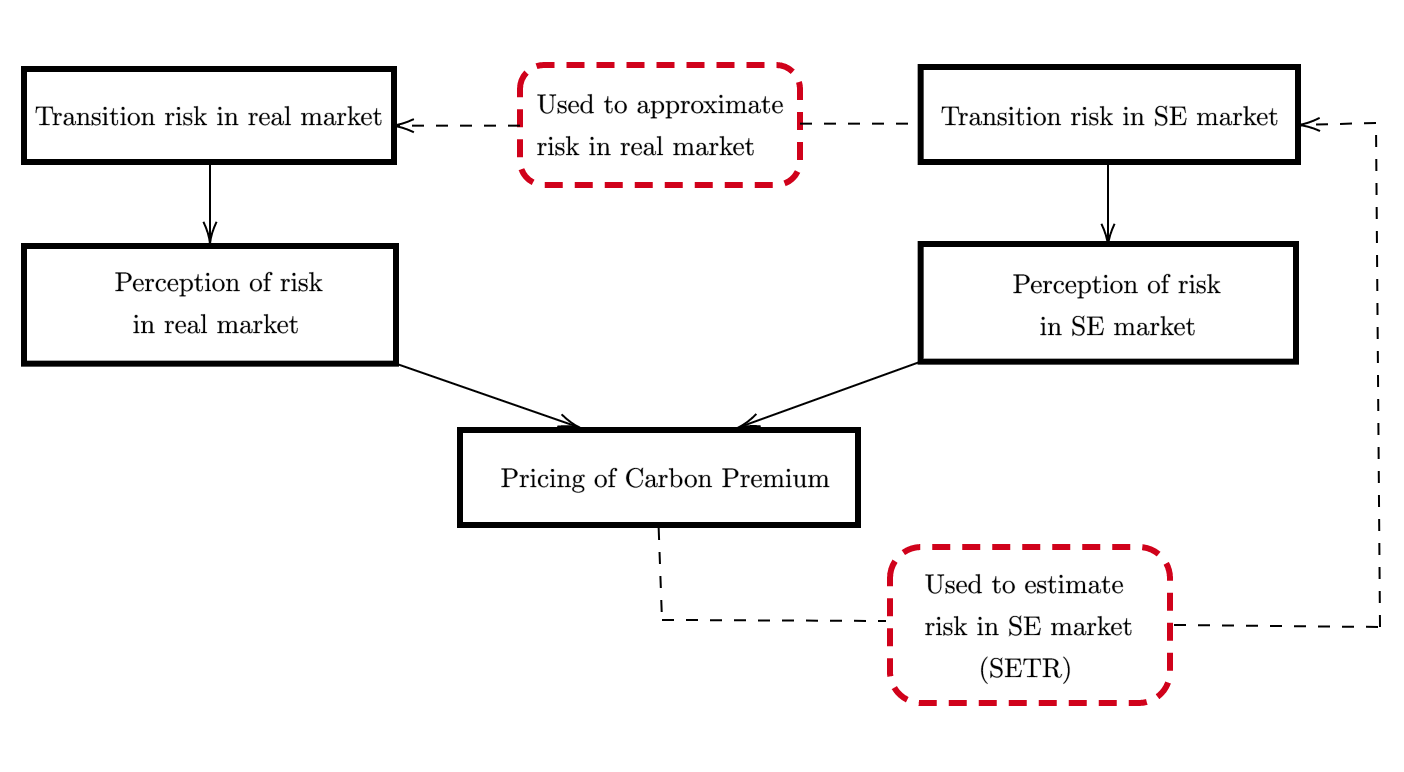"}
\caption{Schematic of estimation of transition risk in the single event framework}
\label{fig:flowchart}
\end{figure}
The SETR is not a real quantity. It is a synthetic measure that we are constructing, in order to get a quantifiable measure for long term carbon exposure. In order to do this, we begin from the definition of a carbon premium. Investors perceive that there is some long term transition risk present in the shares of ``dirty'' stocks. Therefore, a carbon premium is necessary for investors to continue buying and holding the ``dirty'' stocks despite the large risk. In the single event frame work, we construct a thought experiment. We ask what would be the value of long exposure to transition risk on the price of a stock, from a single event transition, that would be offset by the carbon premium that we actually observe in the market. In essence, if the measured value of the carbon premium in real markets is some $p$, we quantify the SETR $\phi(t)$ as the corresponding value of the risk in a hypothetical single event market, which would be offset be the same premium $p$. It would be useful to note that this process is the transpose of how the relationship between premiums and risks are usually determined, where the value of risk is known (or estimated) and the value of the premium is calculated accordingly. However, our methodology does not violate any causal relationship between the risk and the premium. Instead, we consider a hypothetical market where the risk is actually quantifiable, unlike real markets, and ask what the pricing of the risk must have been in the hypothetical single event market in order to generate the same valuation of the carbon premium that we observe in real markets. Since the SETR is calculated for a hypothetical market, it is a ``real'' pricing of carbon transition risk in real markets. Rather, it is a synthetic measure that we argue will be useful to investors as a proxy of the actual exposure, due to no known methods of calculating the ``real'' risk exactly or approximately. In subsequent sections, we will develop the idea of what the hypothetical ``single event market'' is and calculate the SETR in terms of the premium. An example of the price movements of the risk-free price and the corresponding price under the paradigm of carbon premium and SETR has been simulated for different random realizations in Figure \ref{fig:Simulations}.
	
\subsection{Market assumptions}
In order to be able to be able to derive an relationship between the carbon premium and the SETR, we construct a hypothetical single event market where some simplifying assumptions are valid. We expect that some of these assumptions will be realistic even in real markets, while others will not. These assumptions are crucial in reducing a complicated stochastic system to a significantly simpler system with a smaller number of stochastic variables and quantities. The absence of these simplifications would make the principal question too difficult to answer and may even render it unsolvable, thus taking away its utility. We therefore, at present restrict our analysis within the paradigm of these assumptions. These assumptions also call for more literature where the transition risk is calculated for markets with one or more of these assumptions relaxed. 
\begin{assumption}
A non-negative carbon premium exists in the shares of firms with high carbon emissions. When other parameters are kept constant, this premium is greater for firms with greater emissions. 
\end{assumption}
This is a general assumption that we believe is universally valid, based on empirical evidence. 
\begin{assumption}
The financial market is efficient in pricing the carbon premium. 
\end{assumption}
By this we mean that investors in the market are aware of carbon transition risk and the value of the carbon premium with respect to the expected transition risk is correctly priced and universally agreed upon. Apart from assumptions about the carbon risk premium. We also make assumptions about the SETR. 
\begin{assumption}
The entire loss of value of the share of a firm due to low-carbon transition, will be due to a single event that takes place in an instant. 
\end{assumption}
\begin{assumption}
\label{ass-4}
There exists a probability distribution for the arrival of the transition event such that the probability that the transition event occurs in the future is a surety (conditional to an initial time before which the transition has not occurred).
\end{assumption}
Mathematically, the transition event can be modelled by a stochastic risk process $\mathcal{R}(t)$ defined by, 
\begin{equation}
\mathcal{R}(t)=\begin{cases} 
1 &,\ \text{The transition event occurs} \\
0 &,\ \text{The transition event does not occur} \end{cases}
\end{equation}
Let $P[(t_A,t_B)]$ denote the probability of the transition event occurring in some interval $(t_A,t_B)$ \textit{i.e.,} the probability that 
\begin{equation}
\exists \ \ t\in (t_A,t_B)\ \ \text{such that}\ \ \mathcal{R}(t)=1.
\end{equation}
Thus, given initial time $t_0$ such that low-carbon transition is yet to happen, we can model the arrival process by some probability density function $f(t)\in [0,\infty)~\forall~t \in (t_{0},\infty)$ such that, 
\begin{equation}
P[(t_A,t_B)]=\int\limits_{t_A}^{t_B}f(t)~dt\ \ \ \ \ \forall\ \ t_0\le t_A\le t_B<\infty.
\end{equation}
Additionally from Assumption \ref{ass-4} we know that,
\begin{equation}
\int\limits_{t_{0}}^{\infty}f(t)dt=1.
\end{equation} 
Finally, inspired from the pricing of premia in the context of credit risk \cite{bluhm2016introduction}, we make our most crucial assumption. 
\begin{assumption}
\label{ass-no-arb}
The expected surplus return from the carbon premium is equal to the expected loss of value due to the single event transition.
\end{assumption}
This is the assumption on the basis of which we will find the relationship between the carbon premium and the SETR. We shall call this the ``(weak) no-arbitrage assumption'' (as opposed to the strong no-arbitrage condition in Appendix \ref{Appendix-A}).
	
\subsection{Calculation of the SETR}
Consider an investor who buys a stock at time $t_0$. To buy a stock and to hold a position in a stock is equivalent, as an investor will only be interested in profit opportunities in the future. Without loss of generality, we can assume the pre-risk price of the share at time $s=t_{0}$ to be $S(t_{0})=1$. Since the premium is included in the returns, it is also included in the price of the stock $S(s)$ for any general time $s$. We can subsequently define a real valued function $\phi(t):[t_0,\infty)\to \mathbb{R}$ that describes the potential loss of value of the price of each share at any time $t$ at which the transition event occurs. Now, $\phi(t)$ is our estimate of the SETR for an event that occurs at time $t$ \textit{i.e.,} $\mathcal{R}(t)=1$. Since there is only a single transition event, we know that $t$ is unique \textit{i.e.,} if the transition event takes place at time $t$, it will not take place at any other time $s\ne t$.  Therefore we can model the price of a stock generally as, 
\begin{equation}
\mathcal{S}(s)=S(s)-\phi(s)\mathcal{R}(s)\ \ \forall \ \ s\in [t_0,t].
\end{equation}
We can see that this function is equal to $S(s)$ for $s<t$ (until the transition event occurs). 
Based on this, we expect that the existence of the risk premium in the single event market is motivated by the function $\phi(t)$ and the process $\mathcal{R}(t)$ (and therefore implicitly by the density function $f(t)$). Therefore, given $f(t)$, we aim to find the relationship between the risk premium $p$ of the shares of a firm and the SETR function $\phi$.
	
If an investor expects to receive a premium $p(s)$ from time $s=t_{0}$ through $s=t$, then the amount of gains made from the premium till time $t$ is, 
\begin{equation}
A(t)=\int\limits_{t_0}^tp(s)ds.
\end{equation}
Since $t$ is determined by the distribution $f(t)$, the expected earnings from the premium is given by,
\begin{equation}
E(A)=\int\limits_{t_0}^\infty A(t)f(t)dt=\int\limits_{t_0}^\infty f(t)\int\limits_{t_0}^tp(s)ds~dt.
\end{equation}
	
Here, the inner integral gives us the total amount that an investor expects to earn from risk premia above the risk free returns, for a given $t$, while the outer integral averages over all possible times for the transition event with respect to the arrival process $f(t)$. Similarly, we can find the expected value of the SETR as,
\begin{equation}
E(\phi)=\int\limits_{t_0}^\infty\phi(t)f(t)dt.
\end{equation}
Therefore, from Assumption \ref{ass-no-arb} we get, 
\begin{equation}
E(A)=E(\phi).
\end{equation}
Hence, 
\begin{equation}
\label{equation-10}
\int\limits_{t_{0}}^{\infty} f(t)\int\limits_{t_0}^tp(s)ds~dt=\int\limits_{t_{0}}^{\infty}\phi(t)f(t)dt.
\end{equation}
This equation gives us the relationship between the risk premium and the SETR. In order to be able to simplify this further, we shall assume $p(s)=p$ to be a constant for all future times $s>t_0$. Then, 
\begin{equation}
\int\limits_{t_{0}}^{\infty} f(t)\int\limits_{t_0}^tpds~dt=\int\limits_{t_{0}}^{\infty}\phi(t)f(t)dt.
\end{equation}
Therefore, 
\begin{equation}
\int\limits_{t_{0}}^{\infty} f(t)p(t-t_0)dt=\int\limits_{t_{0}}^{\infty}\phi(t)f(t)dt.
\end{equation}
We can split the left hand side into two terms as,
\begin{equation}
p\left[\int\limits_{t_{0}}^{\infty}t_{e}f(t)~dt-t_{0}\int\limits_{t_{0}}^{\infty}f(t)dt\right]=\int\limits_{t_{0}}^{\infty}\phi(t)f(t)dt
\end{equation}
We can see that the first integral on the left hand side is the expected time of arrival of the risk and the second integral is equal to $1$. This is an interesting result because this tells us that the amount an investor can expect to earn from the risk premium does not depend on the form of the arrival process at all and only depends on the expected time of arrival. Therefore, 
\begin{equation}
\label{premium-risk equality}
p\left[E(t)-t_{0}\right]= \int\limits_{t_{0}}^{\infty}\phi(t)f(t)dt.
\end{equation}
	
Furthermore, we see that the expected value of the risk grows linearly with the amount of premium and the expected arrival time of the transition event. This, however is a consequence of assuming a constant expected annual risk premium. A better calculation could use a continuously growing risk premium as determined by Bolton and Kacperczyk (2021), but due to the dearth of available data on risk premiums over sufficiently long periods of time and models that make explicit projections of risk premiums for the future, we are constrained to assume a constant premium.
	
The expected value of the risk however, does not tell us much about the form of $\phi(t)$, but only gives an average. But intuitively, since $\phi$ represents the transition risk, it should be a constant in time, unless the firm makes changes to its operations or scale. This is because the risk is thought of to be the difference in share value between the present value and the value when the firm is forced to cut down its carbon emissions by a certain ``maximum'' amount. Therefore, unless the firm makes changes in the scale of production or the methods of production, the value of the SETR should not change over time. Therefore, we can also assume specifically that $\phi(t)=\varphi$ is a constant. Therefore, 
\begin{equation}
\label{eq-cosnt-SETR}
\varphi=\varphi\int\limits_{t_{0}}^{\infty} f(t)dt= \int\limits_{t_{0}}^{\infty}\varphi f(t)dt.
\end{equation}
Now, substituting equation (\ref{eq-cosnt-SETR}) in equation (\ref{premium-risk equality}), we get,
\begin{equation}
\label{SETR-constant}
\varphi=p\left[E(t)-t_{0}\right]
\end{equation}
This relationship between the risk and premium is extremely important for several reasons. Firstly, because it helps us answer the question that we started out with, namely, to quantify the SETR in the ``single event'' market in terms of the carbon premium. But due to the assumptions of constancy of the premiums and the SETR, the expression is independent of the form of $f$. This is particularly useful for investors, as now quantifying the SETR does not depend on the exact arrival process, but only the expected time of arrival. It does not even depend on the variance of the time of arrival. The expression in equation \ref{SETR-constant} is also the intuitive expectation for the pricing of the SETR under the no-arbitrage conduction, because it produces the result that the value of the SETR is equal to the expected cumulative earnings from the carbon premium for the expected duration for which there will be a premium in the market. 
	
Although this calculation has been performed assuming the premium stays constant into the future, empirical analysis shows that carbon premiums have increased over the years \cite{bolton2021investors}. This can be attributed to the following three scenarios:
\begin{enumerate}[(1)]
\item More investors have become aware about carbon transition risk and the ``average'' premium in the market moves away from $0$ towards the fair price. Thus, it is possible for the estimated premium to increase without the fair pricing of the premium actually increasing. In this case it is a valid assumption to consider a constant premium in the efficient market. 
\item The value of the fair price of the premium increases over time even in the efficient market. In which case a better calculation would involve assuming an increasing form of the premium.
\item A combination of both of the above two factors. In this case as well, the function will probably have an increasing form. 
\end{enumerate}
In the Appendix \ref{Appendix-B} we consider a case where the carbon premium is projected to grow exponentially into the future. This is another realistic assumption as stock prices are conventionally modeled using Geometric Brownian Motions \cite{black2019pricing}, which have exponential solutions. Thus if the premium is maintained at a constant fraction of the share value, it would also likely show an exponential growth (albeit with stochastic variation).

\section{Applications and Future Direction}

Throughout the previous sections, we have introduced the theoretical framework of estimate carbon transition risk via the SETR. We have also calculated the relationship between the risk premium and the SETR under the simple assumptions of a constant premium and a constant risk. For calculations with some relaxations on these assumptions, one may refer to Appendix \ref{Appendix-A} for a calculation where the SETR is not assumed to be a constant, and Appendix \ref{Appendix-B} for a calculation for a geometrically growing premium. In this section, we will discuss the utility of the SETR for market analysts and investors.
	
An obvious benefit of SETR is that there now exists a quantity that can be used as a proxy for the magnitude of the risk. While firm-level carbon premiums are proportional to the SETR, they do not give us information about how the risk scales with the premium. The Single Event framework helps quantify a measure of transition risk that also takes magnitude into account. A consequence of the existence of such a risk measure is that now investors not only know if one asset is riskier than another, they also have a measure of how much riskier an asset is as compared to liquidity. This allows a trader to also compare the risks of holding assets on absolute terms rather than relative terms. Furthermore, this opens avenues of non-arbitrary hedging strategies that allow for finding an optimum investment strategy. The advantage of having an actual risk measure is that investors can now calculate the return-risk trade-off better. One must recognize that the return-risk trade-off in the context of sustainable investing is particularly important due to the presence of a carbon premium and a transition risk which are by themselves measures of returns and risks. Similarly one can use the SETR to develop a dynamic divestment strategy from carbon intensive stocks that allows them to reduce their long term holdings of carbon intensive stocks without losing out on short term earnings from premiums. 
	
The SETR formulation can also be extended to modelling real carbon transition risks associated with legislation such as firm-specific carbon allowances. A transition to an economy where each firm has a limit on the absolute amount of emissions they can produce will lead to several firms having to cut down on their emissions in a very small window, which can be approximated to a single event transition, where the entire transition, from current levels of emissions to the allowed levels happens in a very small interval of time. Thus in a paradigm where carbon allowances or carbon rationing is seen as the prevalent mode of carbon transition, the SETR framework provides a fine approximation of the risk. The premiums can be calculated in terms of the ``abnormal excess returns'' of the share of the firm attributed to the ``excess emissions'' \textit{i.e.,} the difference of current and allowed emissions. The SETR then calculated will give the value of transition risk of the share in the real market. In this case, the real market is also a single event market. Therefore, the SETR calculated under single event assumptions can be extended to real markets.
	
Additionally, the single event framework can be expanded to a $k$-event framework, where there are $k$ possibly non-fatal shocks in the system. A generalization to $k$-events, with $k$ being a random variable will allow the modeling of risk in real markets, rather than using the risk in a single event market as a proxy for the real risk. 

\section{Conclusion}

With the acceleration of climate change, as is the trend, governments will soon have to intervene in the process of firms transitioning to low-carbon methods of production. If firms do not actively transition themselves, they will be forced to see drops in revenue and profitability when governments impose restrictions on emissions. This brings a carbon-transition risk to an investor's portfolio in possibly large magnitudes. However, the magnitude of the risk is difficult to estimate or approximate from market data, leaving investors with rough indicators like absolute emissions of firms and ESG scores. In this article, we develop a novel measure, namely, the SETR, to have an approximate measure for the magnitude of the transition risk. The SETR is thought of as the magnitude of the transition risk in a corresponding ideal, single event market which would yield the same carbon premium as measured from data in real markets. We introduce the ideal assumptions made for the single market and define the relationship between the carbon premium and the SETR. We also explicitly find a simple form for the relationship under the assumption that both the premium and SETR are constants in time. The SETR thus calculated can be used as a proxy for the firm-specific magnitude of carbon transition risk. 
	
\newpage
\bibliographystyle{apalike}
\bibliography{Biblio.bib}
\newpage

\section{Appendix}
\subsection{Calculation of SETR under geometrically growing premiums}
\label{Appendix-B}
Consider a carbon premium that increases with time $s$ according to the following differential equation, 
\begin{equation}
\frac{dp(s)}{ds}=\lambda p(s).
\end{equation}
The solution of this equation gives us,
\begin{equation}
p(s)=p_0e^{\lambda s},
\end{equation}
where $p_0$ is the value of the premium at time $s=t_0$. This model of the premium is based on the assumption that stock prices themselves grow geometrically, and therefore if the premium is maintained at a fixed proportion of the share price, its growth will likely be geometric. Putting this relationship back in equation (\ref{equation-10}), and further assuming a constant SETR, we get,
\begin{equation}
\int\limits_{t_0}^\infty f(t)\int\limits_{t_0}^tp_0e^{\lambda s}ds~ dt= \varphi
\end{equation}
\begin{equation}
\therefore \int\limits_{t_0}^\infty f(t) \frac{p_0}{\lambda} \left[e^{\lambda t}-e^{\lambda t_0} \right] dt =\varphi 
\end{equation}
Hence,
\begin{equation}
\varphi = \int\limits_{t_0}^\infty f(t) \frac{p_0}{\lambda}e^{\lambda t}dt -\frac{p_0e^{\lambda t_0}}{\lambda }
\end{equation}
This gives us the SETR which can be easily calculated with knowledge about $f$, $p_0$ and $\lambda$. 

\subsection{Calculation of SETR under the Strong No-Arbitrage Condition}
\label{Appendix-A}
When we defined the weak no-arbitrage condition before, we defined it as the expected earning from premiums from $s=t_0$ to $s=\infty$ is equal to the expected value of the loss from a change in share value due to a transition at any time between $s=t_0$ and $s=\infty$. However, we can make this condition stronger by enforcing it to remain valid for the expected values up to any time $t'$ rather than only up to $\infty$. 
\begin{assumption}
\label{ass-strong-NA}
The expected surplus return from the carbon premium up to any time $t'$ is equal to the expected loss of value due to the single event transition up to time $t'$, for all $t'\in(t_0,\infty)$.
\end{assumption}
Now, if we assume (and reasonably so) that $\phi(t)$ is a well behaved, bounded function that does not fluctuate wildly, then we can arrive at a form of $\phi(t)$ by making some mathematical manipulations. Accordingly, we study what happens if the transition event does not occur in some time $\Delta t$ after $t_0$. Then,
\begin{equation}
\label{eq-app-one}
p[E(t|t>t_{0}+\Delta t)-t_{0}-\Delta t]=\int\limits_{t_{0}+\Delta t}^{\infty}  \phi(t)Cf(t)dt,
\end{equation}
The validity of this equation is due to the validity of Assumption \ref{ass-strong-NA}. Observing carefully, we can see that this is exactly the relationship of the SETR calculated at some future time $t'$ with an additional factor $C$. Here, the factor $C$ has been added for normalizing the conditional probability distribution such that $t_{e}>t_{0}+\Delta t$, and is equal to the conditional probability of the transition event taking place after $t_{0}+\Delta t$ given that it takes place after $t_{0}$.
Thus,
\begin{equation}
C=\frac{\int\limits_{t_{0}}^{\infty}f(t)dt}{\int\limits_{t_{0}+\Delta t}^{\infty}f(t)dt}=\frac{1}{\int\limits_{t_{0}+\Delta t}^{\infty}f(t)dt}.
\end{equation}
Splitting the integral on the right hand side of equation (\ref{eq-app-one}) into two parts, we get
\begin{equation}
p\left[E(t|t>t_{0}+\Delta t)-t_{0}-\Delta t\right]=C\int\limits_{t_{0}}^{\infty}\phi(t)f(t)dt-C\int\limits_{t_{0}}^{t_{0}+\Delta t}\phi(t)f(t)dt.
\end{equation}
Making the substitution from equation (\ref{premium-risk equality}), we obtain,
\begin{equation}
p\left[E(t|t>t_{0}+\Delta t)-t_{0}-\Delta t\right]=Cp\left[E(t)-t_{0}\right]-C\int\limits_{t_{0}}^{t_{0}+\Delta t}\phi(t)f(t)dt.
\end{equation}
\begin{equation}
\label{eqfourpointtwo}
\therefore \int\limits_{t_{0}}^{t_{0}+\Delta t}\phi(t)f(t)dt=p\left[E(t)-t_{0}\right]-\frac{p}{C}\left[E(t|t>t_{0}+\Delta t)-t_{0}-\Delta t\right].
\end{equation}
Let,
\begin{equation}
t_0+\Delta t=t^{\prime}.
\end{equation}
Therefore, making this substitution in equation (\ref{eqfourpointtwo}), we get, 
\begin{equation}
\therefore \int\limits_{t_0}^{t^{\prime}}\phi(t)f(t)dt=p\left[E(t)-t_{0}\right]-\frac{p}{C}\left[C\int\limits_{t^{\prime}}^{\infty}t~f(t)dt-t^{\prime}\right].
\end{equation}
In order to find the explicit form of $\phi(t)$ we now take the partial derivative of both sides with respect to $t^{\prime}$, to obtain,
\begin{equation}
\frac{\partial}{\partial t^{\prime}}\left(\int\limits_{t_{0}}^{t^{\prime}}\phi(t)~f(t)dt\right)=\frac{\partial}{\partial t^{\prime}}\left[p\{E(t)-t_{0}\}-p\int\limits_{t^{\prime}}^{\infty}t~f(t)dt-t^{\prime}\int\limits_{t^{\prime}}^{\infty}f(t)dt \right]
\end{equation}
\begin{equation}
\therefore \phi(t^{\prime})f(t^{\prime})=p\left\{t^{\prime}f(t^{\prime})+\int\limits_{t^{\prime}}^{\infty}f(t)dt-t^{\prime}f(t^{\prime})\right\}.
\end{equation}
We can therefore write the value of the SETR explicitly in terms of the arrival process $f$, as follows,
\begin{equation}
\label{SETR}
\phi(t^{\prime})=\frac{p}{f(t^{\prime})}\int\limits_{t^{\prime}}^{\infty}f(t)dt.
\end{equation}
In this calculation, we have not used the constant SETR assumption as the constant SETR assumption, under the strong no-arbitrage condition yields only the trivial result that the cumulative premium is the value of the SETR at time $t$ of the transition. 
	
\end{document}